\documentclass[aps,pra,reprint,groupedaddress,amsmath,amssymb,preprintnumbers,showpacs,longbibliography]{revtex4-1}
\usepackage{graphicx}
\usepackage{bm}
\usepackage{dcolumn}
\newcommand{\eq}[1]{Eq.~(\ref{#1})}

\newcommand{\be}{\begin{equation}}
\newcommand{\ee}{\end{equation}}
\newcommand{\ba}{\begin{eqnarray}}
\newcommand{\ea}{\end{eqnarray}}
\newcommand{\bs}{\begin{subequations}}
\newcommand{\es}{\end{subequations}}
\newcommand{\bw}{\begin{widetext}}
\newcommand{\ew}{\end{widetext}}

\begin{document}

\title{Disappearance and reappearance of above-threshold-ionization peaks}

\author{Lars Bojer Madsen}
\affiliation{Department of Physics and Astronomy, Aarhus
University, DK-8000 Aarhus C, Denmark}

\date{\today}

\begin{abstract}
It is shown that above-threshold ionization peaks disappear when the kinetic energy associated with the nondipole radiation-pressure-induced photoelectron momentum in the laser propagation direction becomes  comparable to the photon energy, and how peaks can be made reappear if knowledge of the length and direction of the photoelectron momentum  is at hand and an emission-direction-dependent momentum shift is accounted for. The reported findings should be observable with intense mid-infrared laser pulses.
\end{abstract}

\maketitle

During the last decade a number of experimental strong-field ionization studies have measured and modeled nondipole effects across near-~\cite{Smeenk2011,Haram2019,Doerner2019,Hartung2021,Haram2022} and mid-infrared wavelengths~\cite{Keller2014,Maurer2018,Danek2018,Willenberg2019}. The nondipole signatures are typically in terms of a shift in the maximum of the PMD away from vanishing momentum as would be the expected position of the maximum in the dipole case.  Nondipole effects have also been investigated theoretically at near- and mid-infrared wavelengths in intense pulses using a combination of strong-field approximation (SFA), time-dependent Schr\"odinger equation and tunneling approaches, see, e.g., Refs.~\cite{Keitel2005,Klaiber2013,Yakaboylu2013,Chelkowski2014,Chelkowski2015,Bandrauk2017,He2017,Lein2018JPB,Lein2018PRA,Bandrauk2018,Lein2019PRA,Fritzsche2019,Keitel2019,JensenPRA2020,Ni2020,Lein2021,Lund2021,He2022,Klaiber2022,Madsen2022a,Liang2022,Grafe2022}. The breakdown of the electric dipole approximation in this regime has been know for some time and is due  to radiation pressure and magnetic field effects~\cite{Reiss1990,Reiss2008,Reiss2013,Reiss2014}. 
%In simple terms, the speed of the electron in the laser field is governed by the temporal integral of the electric field, which scales as $A_0= F_0/\omega$. For a fixed $F_0$, the  vector potential $A_0$ increases for decreasing $\omega$ and may attain values that are not vanishing compared to the speed of light, and therefore magnetic Lorentz force effects need be considered.
 Progress in the study of nondipole effects was recently reviewed~\cite{Wang2020,Haram2020,Maurer2021}.

In the photoelectric effect, a peak in the photoelectron spectrum shows up at a kinetic energy $k^2/2= \omega - I_p$, where $\omega$ is the  angular  frequency of the ionizing light and $I_p$ is one of the ionization potentials (atomic units are used throughout). In multiphoton ionization, multiple peaks in the spectrum may show up as described by $k^2/2=n \omega -I_p$, for different integers $n$ making the right-hand side positive, a process known as above-threshold ionization (ATI), where an already free electron absorbs photons. If the ionizing radiation is supplied by an intense laser pulse of sub-picosecond duration,  the ATI spectrum will show peaks at~\cite{Bucksbaum1992,Joachain2011}  
\be
\label{dipoleenergy}
k^2/2 =n \omega -I_p - U_p,
\ee
where the ponderomotive potential, $U_p= F_0^2/(4 \omega^2)$, with $F_0$ the field strength, is the cycle-averaged kinetic energy of the free electron in the laser pulse.  
%These relations can be thought of as expressions of energy conservation, and are as such part of the foundation of the interaction of intense laser light with matter and therefore also described in textbooks~\cite{Joachain2011}. 
Finite pulses have finite bandwidths and result in a broadening of the ATI peaks. 

It is useful to relate \eq{dipoleenergy} to the photoelectron momentum distribution (PMD).  Let the laser propagate in the $x$ direction and be linearly polarized along the $z$ axis. Consider the PMD in the $(k_x, k_z)$ plane. Equation \eqref{dipoleenergy} then expresses that the PMD may attain signal on concentric circles or rings centered at the origin $(k_{x0}, k_{z0}) = (0,0)$ with radii 
\be
\label{radius}
k^\text{D} = \sqrt{2(n \omega -I_p - U_p)}.
\ee
Very recently, however, an experiment~\cite{Lin2022} used a dedicated setup with two counter-propagating 800-nm laser pulses to detect a small  decrease (increase) in the lengths of the momenta in (opposite) the laser propagation direction. Clearly, this finding contradicts the implications of \eq{dipoleenergy}. No matter how intuitively appealing \eq{dipoleenergy} is, it is still an approximation. Equation \eqref{dipoleenergy} is only accurate within the electric dipole approximation for the description of the light-matter coupling, i.e., when the dependence of the coupling on the laser propagation direction is neglected.  Indeed, the small measured shifts in the angle-resolved ATI peaks~\cite{Lin2022} are in agreement with theory predictions~\cite{Reiss1990,Fritzsche2019,JensenPRA2020,Lund2021} that take effects beyond the electric dipole approximation into account. A further exploration of the implications of these effects is the purpose of the present work.

When nondipole effects are considered to first order in $1/c$, with $c$ the speed of light, \eq{dipoleenergy}  is replaced by a relation that takes radiation pressure effects into account. In the $(k_x, k_z)$ momentum plane, energy conservation can be expressed as ~\cite{Reiss1990,Fritzsche2019,JensenPRA2020,Lund2021}
\be
\label{energy}
(k_x - U_p/c)^2/2  + k_z^2/2 = n\omega -I_p - U_p,
\ee 
Equation \eqref{energy} shows that the final momenta are confined to circles with center
\be
\label{center}
(k_{x0}, k_{z0})=(-U_p/c, 0),
\ee
and radius given in \eq{radius}, where in the latter the superscript D indicates that the radius of the ATI ring is as in the dipole case. Therefore, the overall character of the nondipole PMD is determined by signal at energy conserving ATI rings, the centers of which are shifted as specified by \eq{center}. As a consequence of the center shift, the length of the final momenta with respect to the origin in the momentum plane becomes angle dependent. Let $\theta$ denote the polar angle measured with respect to the origin and the positive $z$ axis, i.e., the polarization direction. It is then found that energy conservation is fulfilled  at the momenta
\be
\label{kx}
k_x(\theta)=(k^\text{D} - U_p \sin(\theta)/c) \sin(\theta),
\ee
\be
\label{kz}
k_z(\theta) = (k^\text{D} - U_p \sin(\theta)/c) \cos(\theta).
\ee 

The results above are know~\cite{Lund2021}, and the emission-angle-dependent shift in the length is confirmed by experiment~\cite{Lin2022}, but their implications for intense mid-infrared fields have not been fully realized. Here it will be stressed that the possibility of observing characteristic ATI spectra  with peaks at the energies of the individual photon absorption channels can be significantly  affected by nondipole terms. The possibility of observation or not depends on the magnitude of the radiation-pressure-induced momentum shift, $U_p/c$, in the propagation direction. Clearly, when this shift is of the order of the distance between ATI rings, the  rings are shifted such that integration over emission angle for fixed magnitude of continuum electron energy or fixed momentum with respect to the origin will not capture the $\bm k$ values where energy conservation is fulfilled in the nondipole treatment, \eq{energy}, and the peaks disappear. 

The magnitude of the magnetic nondipole effect under investigation, is conveniently quantified by the parameter 
\be
\label{beta0}
\beta_0= U_p/(2 \omega c),
\ee
 which describes the amplitude in the laser propagation direction of the nondipole figure-8 motion of the electron~\cite{Reiss2014}. When $\beta_0 \simeq 1$, nondipole effects are expected to appear. An alternative measure is obtained when one compares the magnitude of the difference in radii between two consecutive photon-absorption rings to the momentum shift $U_p/c$. Such reasoning leads to the parameter
 \be
 \label{beta1}
 \beta_1= 4 \nu [(U_p^2/c^2/2)/\omega],
 \ee
 where the term in the square bracket is the kinetic energy associated with the radiation-pressure-induced photoelectron momentum in the propagation direction in units of the photon energy, and where the $\nu=5, 6, 7,\dots$ measures the number of photons absorbed above threshold.  The values of $\nu$ start at $5$ to justify a Taylor expansion necessary for obtaining the result in \eq{beta1}.   When the value of $\beta_1$ becomes of the order of unity significant nondipole effects are expected. 

To capture the impact of nondipole effects on ATI spectra obtained by a pulsed laser in as simple physical terms as possible it suffices to consider a finite top-hat pulse with $N$ cycles each of duration $T=2\pi/\omega$. The appearance and disappearance of the ATI peak is govered by the intercycle interference term; see, e.g., Refs.~\cite{Arbo2010,Arbo2012,Maxwell2017,Werby2021} for discussion of such interference within the electric dipole approximation. During each cycle, a phase $e^{ i n 2 \pi \frac{E(\bm k)}{\omega}}$ is picked up by the outgoing electron with an energy including a nondipole correction that depends on the projection,  $k_x$, of $\bm k$ along the propagation direction

\be
\label{E}
E(\bm k) = k^2/2 +I_p + U_p  + k_x U_p/c.
\ee 
The intercycle amplitude following strong-field ionization by an $N$ cycle pulse then reads 
\be
\label{Minter}
M_{\bm k, 0}^\text{Inter} (N)  = \sum_{n=0}^{N-1} e^{ i n 2 \pi \frac{E(\bm k)}{\omega}}.
\ee
Note that the Poisson summation formula, $\sum_{n= - \infty}^{\infty} e^{-2 \pi i n x} = \delta (x-n)$,  in combination with \eq{E}, gives energy deltafunctions and ATI peaks when $N$ goes to infinity, i.e., in this limit one obtains the energy in \eq{energy} and the shift in \eq{center}. 

\begin{figure}
\includegraphics[width=0.45\textwidth]{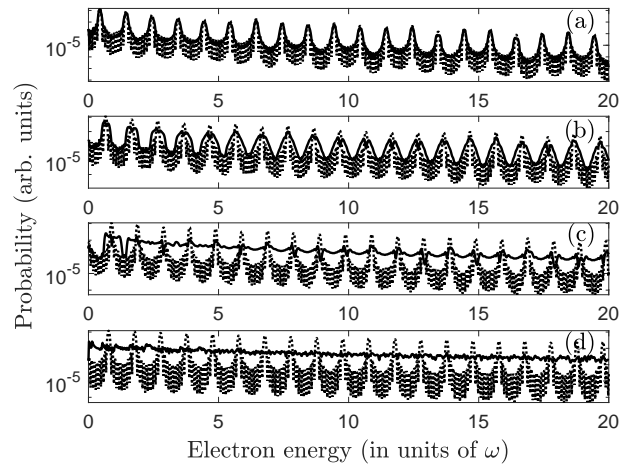}
\caption{Disappearance of ATI peaks. ATI spectra from hydrogen for a 10 cycle pulse with intensity  $10^{14}$ W/cm$^2$  for wavelengths of (a) 1600 nm, (b) 2400 nm, (c) 3200 nm, and (d) 6400 nm. The full curves show results including nondipole effects. The dashed curves show the results within the electric dipole approximation.
}
\label{fig_disappear}
\end{figure}

 In the calculations, which illustrate some consequence of the above discussion for ATI spectra, a typical  intensity of $10^{14}$ W/cm$^2$ is used and $N=10$ cycles is considered for wavelengths $\lambda=$1600 nm, 2400 nm, 3200 nm and 6400 nm. For fixed intensity, the increase in $\lambda$ gives rise to an increase in $U_p \sim \lambda^2$ and therefore a range of $\beta_0$ and $\beta_1$ values [see Table I]. The target is ground state atomic hydrogen. The nondipole strong-field-Hamiltonian approach~\cite{JensenPRA2020} for the SFA ionization amplitude is evaluated in the saddle-point approximation including both inter- and intracyle contributions.  To this end, the approach taken is the one described in Ref.~\cite{Arbo2012} for the dipole case with the nondipole modifications of the phase and the saddle-point solutions described in Ref.~\cite{Madsen2022a}. The PMD is obtained from the norm square of this amplitude. The ATI spectra are obtained by integrating the PMD over electron emission angle for fixed length of the outgoing momentum. Figure \ref{fig_disappear} shows the photoelectron spectra for the considered wavelengths for the nondipole (full curves) and electric dipole (dashed curve) cases. The  figure illustrates that the ATI peaks gradually disappear as $\lambda$ of the driving pulse increases. Indeed, the nondipole ATI spectra at 3200 and 6400 nm are characterized by relatively structureless decreasing curves, the ATI peaks have disappeared. In contrast, the less accurate electric dipole approximation results show clear ATI peaks at all considered $\lambda$'s. Table I collects the values of the parameters $\beta_0$ and $\beta_1$ for the $\lambda$'s considered in Fig.~\ref{fig_disappear}. The values of these parameters support the increase in the nondipole-induced effect seen in Fig.~\ref{fig_disappear} in the sense that a smearing out of the peaks occurs when they attain values around  unity.

% \begin{table}
% \caption{\label{molparam} HOMO energies ($E_\text{HOMO}$) and  molecular-fixed frame polarizability components ($\alpha_{xx}$, $\alpha_{zz}$) of the O$_2^+$, CO$_2^+$, and CS$_2^+$ cations as obtained from quantum chemistry calculations at the B3LYP level of theory (MP2 for CO$_2^+$ polarizability). All quantities are in atomic units.}
% \begin{ruledtabular}
%\begin{tabular}{lccc}
%&$E_\text{HOMO}$ & $\alpha_{xx}$ & $\alpha_{zz}$ \\
%\hline
%O$_2^+$  & -0.3199 &  5.08 &  9.41  \\
%CO$_2^+$ & -0.3845 &  7.29 & 30.42  \\
%CS$_2^+$ & -0.2805 & 28.32 & 76.77  \\
%\end{tabular}
%\end{ruledtabular}
%\end{table}

 \begin{table}
 \caption{\label{table} Parameters $\beta_0$ [\eq{beta0}] and $\beta_1$ [\eq{beta1}] used to assess the importance of nondipole effects for an intensity of $10^{14}$ W/cm$^2$ for the wavelengths considered in Fig.~1.}
 \begin{ruledtabular}
\begin{tabular}{lcc}
Wavelength, $\lambda$ & $\beta_0$ & $\beta_1(\nu=5)$ \\
\hline
1600 nm & 0.11 &  0.014  \\
2400 nm & 0.37 &  0.11  \\
3200 nm & 0.90 &  0.46  \\
6400 nm & 7.18 & 14.7  \\
\end{tabular}
\end{ruledtabular}
\end{table}

\begin{figure}
\includegraphics[width=0.45\textwidth]{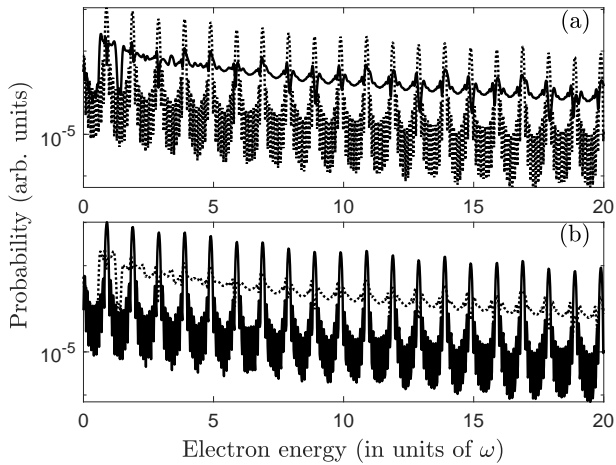}
\caption{Reappearence of ATI peaks. (a) ATI spectra from hydrogen for a 10 cycle pulse with intensity  $10^{14}$ W/cm$^2$  and $\lambda=3200$ nm. The full curves show results  including nondipole effects. The dashed curves show results within the electric dipole approximation. (b) As (a), but with momenta shifted according to Eqs.~\eqref{kx}-\eqref{kz} to account for the nondipole shift of the center of the energy conserving ATI rings according to \eq{center}.}
\label{fig_reappear}
\end{figure}

In the analysis of PMD data, ATI peaks at energies of the different photon absorption channels can be made to reappear if one integrates over the angle of the outgoing electron not with respect to the origin $(k_{x0}, k_{z0})=(0,0)$, but with respect the shifted center $(k_{x0}, k_{z0})=(-U_p/c, 0)$. The accompanying shifts of the momenta as given by Eqs.~\eqref{kx}-\eqref{kz}  guarantee that the concentric energy conserving rings refer to the shifted center. Figure \ref{fig_reappear} shows the result for the ATI spectra at 3200 nm when the integration is done with respect to (a) $(0,0)$  and (b)  $(-U_p/c, 0)$. After the appropriate emission-angle-dependent modification of the lengths of the outgoing momenta, the data in (b) show that ATI peaks reappear in the nondipole data, while the peak structure in the dipole ATI spectrum is washed out by the shifts in Eqs.~\eqref{kx}-\eqref{kz}.

Note that the parameters for the laser used for the results in Fig.~\ref{fig_reappear} are similar to those reported in some experiments~\cite{Keller2014,Maurer2018,Danek2018,Willenberg2019}. An experimental exploration of the present nondipole effects  in this wavelength regime therefore seems possible.  It is noted that ATI peaks were not clearly observed in the mentioned experiments~\cite{Keller2014,Maurer2018,Danek2018,Willenberg2019}. Their absence is consistent with the present findings. Prior to this work, a plausible reason for the absence of ATI peaks could be constructed in a picture involving a field-driven, rather than a multiphoton-driven ionization mechanism. While the field driven aspect is certainly important, it is beyond doubt that the quantized nature of the photon energy should show up as soon as there are more than a few cycles in the pulse. The result in Fig.~\ref{fig_reappear} shows how the peaks may be made reappear by appropriate analysis of momentum resolved data.

 It is likewise noted that the shifts of the ATI peaks in the experimental 800-nm work~\cite{Lin2022} are small and the peaks are still observed; again consistent with the present findings and the small parameter values $\beta_0\simeq 1.4 \times 10^{-2}$, $\beta_1(\nu=5)  \simeq 4.5 \times 10^{-4}$ under those experimental conditions.  Related to experiment, one is  led to a general consideration of whether or not it is possible to analyze experimental data such that ATI peaks can be made reappear. Indeed, if a nondipole PMD is available, the peak structure can be revealed by accounting for the $\theta$-dependent shifts in length of the individual photon absorption channels as described by Eqs.~(\ref{kx})-(\ref{kz}). This reconstruction requires measurement of momenta as vectors. Such vectorial information can be obtained, e.g., in a COLTRIMS apparatus [42, 43]. If, on the other hand, the ATI spectrum is obtained without knowledge of the emission direction, reconstruction of momenta with respect to the shifted center of the ATI rings can not be performed and the ATI peaks may disappear for intense mid-infrared laser pulses.

It is of interest to note that the measure $\beta_1$ depends on the number of photons $\nu$ absorbed above threshold. The right-hand side \eq{beta1}  increases linearly with $\nu$. This can be exploited to observe nondipole effects at the higher-order ATI peaks even at near-infrared wavelengths. For example at 800 nm and for an intensity of $5 \times 10^{14}$ W/cm$^2$ a value of $\beta_1 = 1$ is obtained for $\nu \simeq 89$ well below the classical cut-off of $10 U_p/\omega \simeq 192$. These kind of predictions of trends have been confirmed by calculating spectra for a range of parameters (not shown). 
%Note that at $10^{14}$ W/cm$^2$,  the laser-induced velocity $A_0$ is much smaller than the speed of light, $A_0/c \ll 1$ for all wavelengths considered making the expansion to first order in $1/c$ accurate.

%
%Experiemtn with mid-nfrared from free-electron source
%\cite{Vrakking2008,Vrakking2012,Huismans2011,Vrakking2013} $U_p/(\omega c) \ll 1$
%

The present work analyzed nondipole ATI spectra. Due to the presence of the nondipole term $k_x U_p/c$ proportional to $k_x$ along the laser propagation direction in the nondipole continuum energy of the outgoing electron, the PMD shift by $-U_p/c$ in the propagation direction, i.e., by $U_p/c$ opposite to the propagation direction.  The intercycle contribution determines the $\bm k$ points with constructive interference and lead to energy conserving rings in the limit of infinite many cycles.  For intense mid-infrared laser pulses, it was shown how the expected peaks in the ATI spectrum may disappear due to nondipole effects, but can be brought to reappear when the  nondipole-induced shift of the center of the ATI rings is taken into account in the analysis of the photoelectron momenta. These emission-angle-dependent momentum modifications are useful in the analysis of experimental data with intense mid-infrared lasers, and allow a recovery of  clear ATI peaks.

%\begin{acknowledgments}
This work was supported by the Independent Research Fund Denmark (Grant No. 9040-00001B and 1026-00040B).
%\end{acknowledgments}

%\bibliography{referencer}


%\newpage

\end{document}